# Short-wavelength secondary instabilities in homogeneous and stratified shear flows

## H. M. Aravind[1], Manikandan Mathur[1]† and Thomas Dubos[2]

[1]Department of Aerospace Engineering, Indian Institute of Technology Madras, Chennai - 600036, India

[2]Laboratoire Météorologie Dynamique, École Polytechnique, 91120 Palaiseau, France

We present a numerical investigation of three-dimensional, short-wavelength linear instabilities in Kelvin-Helmholtz vortices in homogeneous and stratified environments. The shear instability of a parallel shear flow, whose velocity and buoyancy profiles define the Reynolds number $Re$ and Richardson number $Ri$, is simulated using two-dimensional direct numerical simulations. The resulting time-dependent vortices are used as base flows on which a local stability analysis is performed, assuming the base flow to be quasi-steady. For the homogeneous case of $(Re, Ri) = (300, 10^{-8})$, the elliptic instability at the vortex core dominates at early times, before being taken over by the hyperbolic instability at the vortex edge. For the stratified case of $(Re, Ri) = (300, 0.08)$, the early time instabilities comprise a dominant elliptic instability at the core and a hyperbolic instability strongly influenced by stratification at the vortex edge. At intermediate times, the local approach shows a new branch of instability (convective branch) that emerges at the vortex core and subsequently moves towards the vortex edge. A few more convective instability branches appear at the vortex core and move away, before coalescing to form the most unstable region inside the vortex periphery at large times. The dominant instability characteristics from the local approach are shown to be in good qualitative agreement with results based on global instability studies for both homogeneous and stratified cases. Compartmentalized analyses, where the buoyancy or velocity gradient terms are omitted from the local stability equations, are then used to elucidate the role of shear and stratification on the various identified instabilities. The role of buoyancy is shown to be critical after the primary Kelvin-Helmholtz instability saturates, with the dominant convective instability shown to occur in regions with the strongest statically unstable layering. We conclude by highlighting the potentially insightful role that the local approach may offer in understanding the secondary instabilities in other numerically simulated vortical flows.

## 1. Introduction

Vortex instabilities are now recognized to be fundamental in understanding various phenomena in natural and engineering flows. For example, complex three-dimensional structures resulting from vortex instabilities often play an important role in the transition to turbulence; coherent vortical motions and associated dynamics continue to persist in the turbulent regime as well (Pullin & Saffman 1998). Several previous studies (Leibovich 1978; Kerswell 2002; van Heijst & Clercx 2009) have therefore addressed various aspects of vortex instabilities using laboratory experiments, numerical simulations and stability analyses. In this paper, we perform a local stability analysis of the vortices that result

† Email address for correspondence: manims@ae.iitm.ac.in

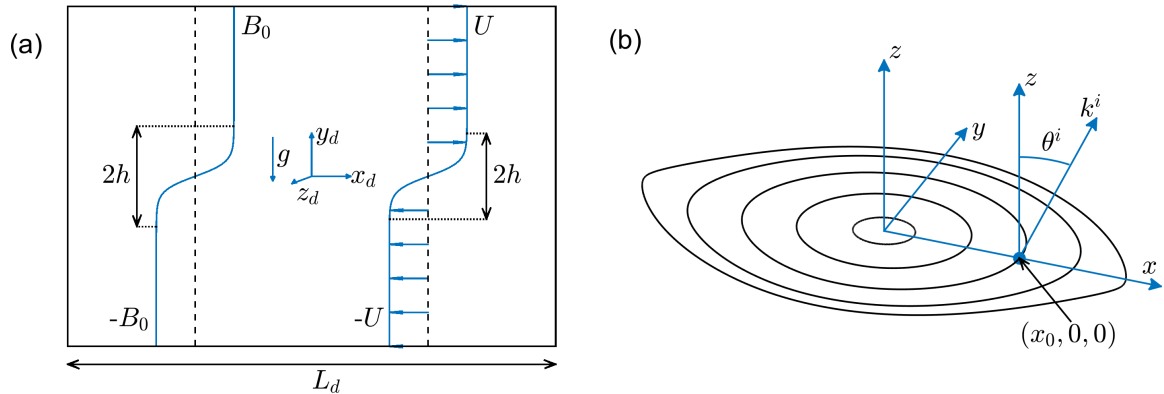

Figure 1: (a) A schematic of the flow domain and initial conditions (buoyancy $B_d^i$ and velocity $u_d^i$ on the left and right, respectively, with $B_0 = N^2 h$) used for two-dimensional numerical simulation of Kelvin-Helmholtz vortices. $L_d$ ($= 4\pi h$) is the length of the computational domain. $2h$ represents the width of both the shear layer and buoyancy layer. (b) A few numerically extracted closed streamlines in the KH vortex on the $xy$−plane. An initial perturbation wave vector $\boldsymbol{k}^i$, making an angle $\theta^i$ with the $z$-axis, of a three-dimensional perturbation that evolves on one of the extracted streamlines is also shown.

from the Kelvin-Helmholtz (KH) instability, in both homogeneous and stably stratified shear flows.

The KH instability manifests in plane shear flows that contain an inflection point in the one-dimensional velocity profile. In the presence of stable stratification, often encountered in the atmosphere and the ocean, the KH instability occurs if the stratification is sufficiently weak. However, if $Ri > 0.25$ is satisfied everywhere in a parallel, stratified, inviscid flow, no linear instability is possible (Miles 1961; Howard 1961). Here, the Richardson number $Ri$ is a measure of the ratio between the stratification and shear effects. The primary KH instabilities that occur for $Ri < 0.25$ result in the formation of an array of vortices (Thorpe 1973) that are connected by braid-like regions, with the resulting flow characterized by the presence of elliptic and hyperbolic points. The focus of the current study is a three-dimensional stability analysis of these two-dimensional flows that result from a primary KH instability, i.e. secondary instabilities in homogeneous and stably stratified shear flows.

Extensive global mode linear stability analyses, along with energy budget calculations, have been reported for the two-dimensional KH vortices that form in stratified shear flows (Klaassen & Peltier 1985, 1991; Caulfield & Peltier 2000); all these studies make a quasi-steady assumption for the base flow and study the temporal evolution of the secondary instability characteristics. Klaassen & Peltier (1985) performed a global mode stability analysis on the numerically generated two-dimensional base flows that result from the primary KH instability for fixed parameter values of $Re = 500$ ($Re$ is the Reynolds number associated with the initial one-dimensional shear flow) and $Ri = 0.07$. They report three-dimensional instabilities that are convective in nature, with the corresponding eigenmodes focused in the statically unstable regions of the base flow. Klaassen & Peltier (1991) extended the study of Klaassen & Peltier (1985) to investigate the effects of Richardson number on the three-dimensional secondary instabilities, but at $Re = 300$. They conclude that the base flow shear drives the secondary instabilities at early times for all $Ri$. In contrast, the secondary instabilities at large times derive their energy from convective overturning in the vortex centres for Richardson numbers in the range $0.065 \leqslant Ri \leqslant 0.13$. In summary, global mode analyses have revealed an elliptic secondary instability at the centre and a dominant hyperbolic instability at the vortex edge in homogeneous shear flows. In stratified flows, the central elliptic instability along with a more dominant convective instability near the periphery of the vortex have been reported. In this paper, we perform a local stability analysis to complement the results from previous global mode approaches, particularly in terms of identifying specific regions of various secondary instabilities and associated mechanisms.

The local stability approach (Lifschitz & Hameiri 1991) employs the WKB approximation to investigate three-dimensional, short-wavelength instabilities in a given base flow. The local stability equations, governing the evolution of leading order perturbation amplitudes on fluid trajectories in the base flow, have previously been used to investigate various instability mechanisms in several idealized vortex models. The local approach has provided significant insight into the effect of various factors like strain, background rotation, stratification and axial flow on vortex models such as Stuart vortices, Taylor-Green vortices and a Rankine vortex (Miyazaki & Fukumoto 1992; Miyazaki 1993; Dizès & Eloy 1999; Godeferd *et al.* 2001; Mathur *et al.* 2014; Nagarathinam *et al.* 2015). Being computationally inexpensive, the local approach has helped identify centrifugal, elliptic and hyperbolic instabilities on specific streamlines in the strongly non-parallel model vortex flows. The local approach has also been used on numerically simulated two-dimensional wake flows (Gallaire *et al.* 2007; Giannetti 2015; Citro *et al.* 2015; Jethani *et al.* 2017), but never on a numerically simulated base flow in which stratification plays an important role.

Stuart vortices have often been used to model mixing layer vortices in homogeneous flows. Klaassen & Peltier (1991); Rogers & Moser (1992), however, report that Stuart vortices may not capture all the secondary instability characteristics in homogeneous mixing layers. Furthermore, the relevance of the Stuart vortices model to describe mixing layer vortices in stratified environment is also unclear. In this paper, we perform local stability calculations on KH vortices simulated using two-dimensional numerical simulations, thus eliminating the approximations associated with idealized vortex models. Three-dimensional numerical simulations (Metcalfe *et al.* 1987; Staquet & Riley 1989; Rogers & Moser 1992; Caulfield & Peltier 2000) and laboratory experiments (Thorpe 1987) have revealed the emergence of small-scale coherent structures in homogeneous and stratified mixing layer vortices. Global stability analysis (Caulfield & Peltier 2000) has also reported on large wavenumber instabilities in both homogeneous and stratified cases, thus suggesting that the secondary instabilities may be amenable to the short-wavelength approximation that the local approach assumes. In § 2, we present the details of two-dimensional numerical simulations used to generate the base flows, the local stability equations and the methods adopted to compute growth rates associated with various instabilities. § 3 presents the results for the homogeneous (unstratified) base flow, followed by a detailed investigation of a representative stratified scenario. Physical interpretation of our results are provided at the end of § 3, followed by our discussion and conclusions in § 4.

## 2. Theory and Methods

We study three-dimensional, short-wavelength instabilities on the two-dimensional vortical flow that develops upon numerically solving non-dimensional forms of the mass, momentum and buoyancy equations in the limit of the Boussinesq approximation (Mkhinini

*et al.* 2013):

$$\boldsymbol{\nabla} \cdot \boldsymbol{u} = 0, \tag{2.1}$$

$$\frac{\partial \boldsymbol{u}}{\partial t} + (\boldsymbol{u} \cdot \boldsymbol{\nabla})\boldsymbol{u} + \boldsymbol{\nabla} p = \frac{1}{Re}\nabla^2 \boldsymbol{u} + RiB\boldsymbol{e}_y, \tag{2.2}$$

$$\frac{\partial B}{\partial t} + (\boldsymbol{u} \cdot \boldsymbol{\nabla})B = \frac{1}{RePr}\nabla^2 B, \tag{2.3}$$

on a two-dimensional Cartesian grid $(x, y) = (x_d/h, y_d/h) \in [-2\pi, 2\pi] \times (-\infty, \infty)$ (see figure 1a for a schematic) with initial conditions (denoted by superscript $i$):

$$\boldsymbol{u}^i = \frac{\boldsymbol{u}^i_d}{U} = \tanh(y)\boldsymbol{e}_x, \quad B^i = \frac{B^i_d}{N^2 h} = \tanh(y). \tag{2.4}$$

The various quantities — spatial coordinates, time, velocity, pressure & buoyancy — with and without the subscript $d$ are dimensional and non-dimensional, respectively. The dimensional buoyancy is defined as $B_d = g(1 - \rho_d/\rho_{ref})$, where $\rho_d$ and $\rho_{ref}$ are the dimensional density field and constant reference density and $g$ is the magnitude of the acceleration due to gravity that acts along negative $\boldsymbol{e}_y$. Spatial coordinates and time have been non-dimensionalized by the shear layer half-width $h$ and the advective time scale $h/U$, respectively, and $\boldsymbol{e}_x$ & $\boldsymbol{e}_y$ are the unit vectors along $x$ & $y$. Buoyancy and pressure are non-dimensionalized by $N^2 h$ and $\rho_{ref}U^2$, respectively, where $N$ is the Brunt Vaisala frequency at $y = 0$ in the initial condition (equation 2.4). The non-dimensional parameters that govern the flow dynamics are the Reynolds number $Re \ (= Uh/\nu)$, the Richardson number $Ri \ (= N^2/(U/h)^2)$ and the Prandtl number $Pr \ (= \nu/\kappa)$, where $\nu$ and $\kappa$ are the kinematic viscosity and the buoyancy diffusivity, respectively.

We choose the non-dimensional parameters to be $(Re, Pr) = (300, 1)$ with $Ri = 10^{-8}$ and 0.08 respectively for the homogeneous and stratified cases. These specific values allow us to make comparisons with the global analysis of Klaassen & Peltier (1991). For numerical resolution, the vertical domain $y \in (-\infty, \infty)$ is mapped on to a finite domain using a *tanh* transformation, and the flow field is assumed to be unperturbed at $y \to -\infty$ and $y \to +\infty$ (Mkhinini *et al.* 2013). Setting $L_d = 4\pi h$, which is approximately the wavelength of the most unstable primary wave, along with periodic boundary conditions in the horizontal allows us to simulate the evolution of one coherent vortex that forms as a result of the primary KH instability. The vorticity field for the homogeneous case at $Re = 300$, as calculated by numerical simulations, at three different times are shown in figures 2(a)-(c). While the horizontal size of the vortex hardly changes from $t = 40$ to $t = 100$, the vorticity field gets redistributed such that the vertical extent of the vortex increases with time till $t \approx 70$, after which the vortex saturates in its size and shape. We consider these simulated flow fields to be frozen at each instant and use them as steady base flows for our stability analysis.

For the local stability analysis (Lifschitz & Hameiri 1991), we consider a decomposition of the velocity, pressure and buoyancy fields into a sum of base flow (denoted by subscript $B$) and perturbation (denoted by prime) fields as $\boldsymbol{u} = \boldsymbol{u}_B + \boldsymbol{u}', p = p_B + p', B = b_B + b'$, and substitute in the governing equations (2.1)-(2.3) to derive the linearized governing equations for the perturbations. The perturbation fields, within the limits of the WKBJ approximation, are written as:

$$\{\boldsymbol{u}', p', b'\} = \exp\left(\frac{i\phi(\boldsymbol{x}, t)}{\epsilon}\right)[\{\boldsymbol{a}(\boldsymbol{x}, t), \pi(\boldsymbol{x}, t), b(\boldsymbol{x}, t)\} + \epsilon\{\boldsymbol{a}_\epsilon(\boldsymbol{x}, t), \pi_\epsilon(\boldsymbol{x}, t), b_\epsilon(\boldsymbol{x}, t)\} + \ldots], \tag{2.5}$$

where $\epsilon \ll 1$ is a small parameter, and hence indicative of the perturbations being of short wavelength. The perturbation wave vector is given by $\boldsymbol{k} = \boldsymbol{\nabla}\phi/\epsilon$, where $\phi$ is a real-valued scalar function. Substituting the solution forms in (2.5) into the inviscid (no diffusion in both momentum and buoyancy) equations governing small-amplitude perturbations, and retaining only the $O(\epsilon^{-1})$ and $O(\epsilon^0)$ terms give the local stability equations that govern the evolution of the wave vector and the leading order perturbation amplitudes (Miyazaki & Fukumoto 1992):

$$\frac{\mathrm{d}\boldsymbol{k}}{\mathrm{d}t} = -(\nabla \boldsymbol{u_B})^T \cdot \boldsymbol{k}, \tag{2.6}$$

$$\frac{\mathrm{d}\boldsymbol{a}}{\mathrm{d}t} = -(\nabla \boldsymbol{u_B}) \cdot \boldsymbol{a} + Rib\boldsymbol{e}_y + \frac{\boldsymbol{k}}{|\boldsymbol{k}|^2}(2((\nabla \boldsymbol{u_B}) \cdot \boldsymbol{a}) \cdot \boldsymbol{k} - Rib\boldsymbol{e}_y \cdot \boldsymbol{k}), \tag{2.7}$$

$$\frac{\mathrm{d}b}{\mathrm{d}t} = -\boldsymbol{a} \cdot \boldsymbol{\nabla} b_B, \tag{2.8}$$

along with the constraint of $\boldsymbol{a} \cdot \boldsymbol{k} = 0$ that comes from the continuity equation. Here, $\mathrm{d}/\mathrm{d}t = \partial/\partial t + \boldsymbol{u}_B \cdot \boldsymbol{\nabla}$ is the material time derivative in the base flow, i.e. equations (2.6)-(2.8) represent the evolution of $\boldsymbol{k}$, $\boldsymbol{a}$ & $b$ along fluid particle trajectories in the base flow $\boldsymbol{u}_B$.

In our study, we consider instantaneous snapshots of the numerically simulated two-dimensional base flows to represent steady base flows in equations (2.6)-(2.8). Furthermore, we restrict our stability calculations to closed streamlines in the base flow. For every closed streamline, we consider only those wave vectors that are periodic upon integrating equation (2.6) once around the closed streamline. Such periodic wave vectors are given by (Mathur *et al.* 2014):

$$\boldsymbol{k} = \beta\boldsymbol{\nabla}\psi + \cos\theta^i \boldsymbol{e}_z, \tag{2.9}$$

where $\psi(x,y)$ is the stream function describing the base flow through the relation $\boldsymbol{u}_B = (-\partial\psi/\partial y)\boldsymbol{e}_x + (\partial\psi/\partial x)\boldsymbol{e}_y$. Owing to the linearity of the equations (2.6)-(2.8) with respect to $\boldsymbol{k}$, we restrict our calculations to $|\boldsymbol{k}^i| = 1$, and hence choose $\beta = \sqrt{(1-\cos^2\theta^i)/|\boldsymbol{\nabla}\psi^i|^2}$, with $\theta^i$ representing the angle made by the initial wave vector $\boldsymbol{k}^i$ with the $z$-axis (depicted in figure 1b). For a given closed streamline, which is fixed by the initial condition $(x_0, 0, 0)$ with $x_0 > 0$ (shown in figure 1b), we solve the perturbation amplitude evolution equations (2.7)-(2.8) for 1000 different values of $\theta^i$ in the range $[0, \pi/2]$.

For each streamline (uniquely identified by the initial condition $x_0$ on the $x$-axis) and a given $\theta^i$, equations (2.7)-(2.8) were integrated numerically using Runge-Kutta fourth order scheme from 0 to $T$ for four different initial conditions: $\boldsymbol{a}_1^i = [1,0,0,0]$, $\boldsymbol{a}_2^i = [0,1,0,0]$, $\boldsymbol{a}_3^i = [0,0,1,0]$ and $\boldsymbol{a}_4^i = [0,0,0,1]$, where $\boldsymbol{a}_j^i = [a_x^i, a_y^i, a_z^i, b^i]$, to obtain the final amplitude vectors $\boldsymbol{a}_1^f$, $\boldsymbol{a}_2^f$, $\boldsymbol{a}_3^f$ and $\boldsymbol{a}_4^f$. Here, $T$ is the time period over which a fluid particle goes around the closed streamline once. The growth rate, using results from Floquet theory (Chicone 2000), is then computed as:

$$\sigma(x_0, \theta^i) = (1/T)\max[Re(log(\lambda_j))], \tag{2.10}$$

where $\lambda_j (1 \leqslant j \leqslant 4)$ are the eigenvalues of the $4\times 4$ matrix $\boldsymbol{M} = [(\boldsymbol{a}_1^f)^\intercal, (\boldsymbol{a}_2^f)^\intercal, (\boldsymbol{a}_3^f)^\intercal, (\boldsymbol{a}_4^f)^\intercal]$. For the integration of equations (2.7)-(2.8), we discretize the time period $T$ by 4000 equispaced time intervals.

The assumption of a steady base flow for growth rate computations is valid only if the time scale associated with the growth rate is much smaller than the time scale associated with the evolution of the numerically simulated base flow. Such a condition is reasonably satisfied if $\sigma > \sigma_{KH}$, where $\sigma$ is the calculated growth rate for the perturbations that grow on the simulated base flow, and $\sigma_{KH}$ the growth rate associated with the evolution of the simulated primary vortex. Here, we compute $\sigma_{KH}$ based on the evolution of the kinetic energy associated with the flow deviation from the initial conditions, as given in equation (3.7) of Klaassen & Peltier (1991). The extent of validity of the condition $\sigma > \sigma_{KH}$ is presented wherever appropriate in § 3.

Previous studies using the local stability approach have identified centrifugal, elliptic and hyperbolic instabilities in various idealized vortex models (Godeferd *et al.* 2001). The centrifugal instability on a given streamline is often associated with the most unstable wave vector being purely spanwise ($\theta = \theta^i = 0$). In contrast, the most unstable wave vector associated with the elliptic instability makes an angle of around $\pi/3$ with the spanwise direction (Kerswell 2002). The hyperbolic instability (Leblanc 1991), which occurs on streamlines that pass through regions in the neighbourhood of hyperbolic points, is also characterized by purely spanwise perturbations being most unstable. It is, however, important to note that these classical signatures of various instabilities can be significantly modified in the presence of factors such as stratification (Miyazaki & Fukumoto 1992; Miyazaki 1993), background rotation (Godeferd *et al.* 2001) and/or axial flow (Mathur *et al.* 2014; Nagarathinam *et al.* 2015), and are hence used as guidelines rather than strict criteria.

## 3. Results

### 3.1. *Homogeneous flow - $(Re, Ri) = (300, 10^{-8})$*

For the homogeneous (simulated by assigning $Ri = 10^{-8}$ in the base flow simulations) case at $Re = 300$, the base flows were generated until $t = 200$ for every integer time $t$. The base flow vorticity field at three representative times are shown in figures 2(a)-(c). At every $t$, 100 different closed streamlines were extracted, with $x_0 = 0$ denoting the vortex centre. The maximum $x_0 (= x_{0_l})$ denotes the closed streamline furthest from the centre and passing through regions close to the hyperbolic points at $(-2\pi, 0)$ and $(2\pi, 0)$. Figures 2(d)-(f) show the variation of the computed growth rate $\sigma$ (equation 2.10) as a function of $x_0$ on the x-axis and the initial wave vector angle $\theta^i$ on the y-axis at three different times: $t = 40$, 60 and 100. At $t = 40$ (figure 2d), the vortex centre ($x_0 = 0$) is evidently unstable, with the corresponding growth rate attaining a maximum of 0.147 at $\theta^i = 51.4°$, a value close to $\pi/3$. It is well known that elliptic instability is characterized by the corresponding most unstable wave vector occurring at $\theta = \pi/3$ (Bayly 1986; Kerswell 2002), owing to which we conclude that the vortex core at $t = 40$ is susceptible to inviscid elliptic instability. The streamlines in the immediate neighbourhood of the vortex centre, which are not necessarily exactly elliptic, are observed to be unstable in the range of $26.7° \leqslant \theta^i \leqslant 74.1°$. The range of unstable $\theta^i$ increases as we go away from the vortex centre, with the corresponding most unstable $\theta^i$ moving towards $\theta^i = 0$ as we approach the edge of the vortex. The instability at the edge of the vortex is hyperbolic, as signified by the presence of a hyperbolic point in its immediate neighbourhood and the most unstable $\theta^i$ occurring close to $\theta^i = 0$. We recall from earlier studies (Leblanc 1991; Godeferd *et al.* 2001) that hyperbolic instability, in general, is most severe for purely spanwise perturbations, i.e. $\theta = 0$. In summary, at $t = 40$, the vortex core and the edge are susceptible to elliptic and hyperbolic instabilities, respectively, and the intermediate streamlines exhibit a combination of both. As indicated by the location of maximum $\sigma$ over the entire $x_0 - \theta^i$ plane (red circle in figure 2d), the elliptic instability at the core dominates the hyperbolic instability at the edge at $t = 40$.

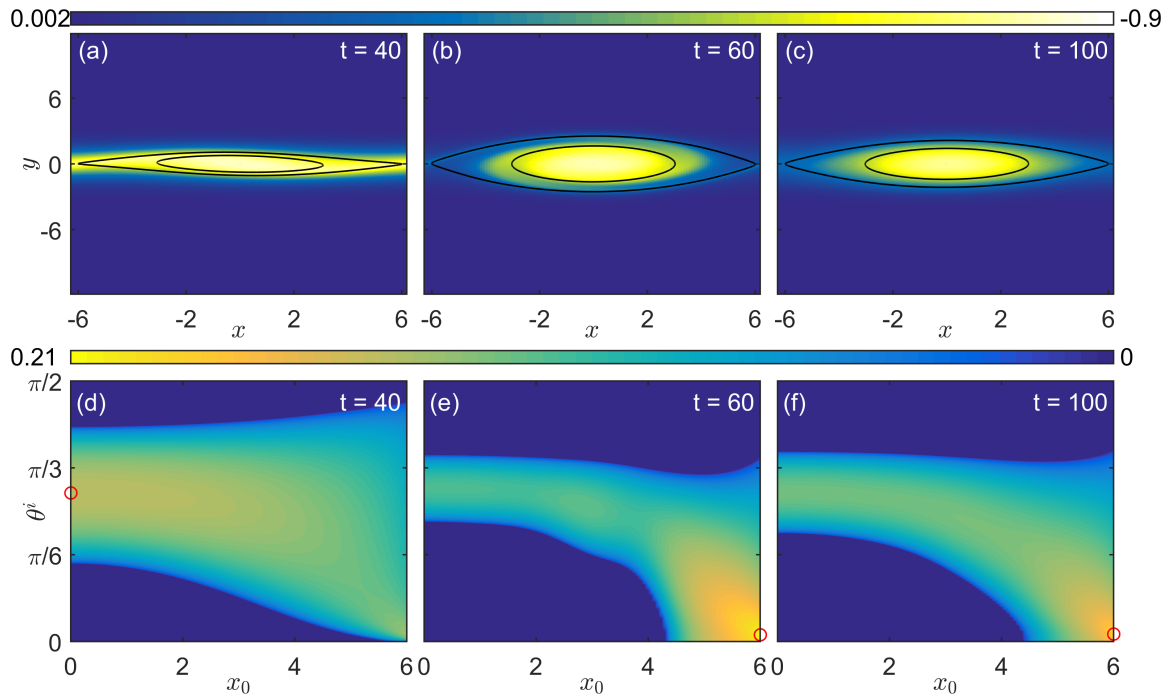

Figure 2: (Top row) Vorticity field, obtained using two-dimensional numerical simulations, at three different times: (a) $t = 40$, (b) $t = 60$ and (c) $t = 100$ for the homogeneous case of $(Re, Ri) = (300, 10^{-8})$ (§ 3.1) along with two representative numerically extracted closed streamlines. (Bottom row) Instability characteristics for the homogeneous case: growth rate ($\sigma$) computed using the local stability framework as a function of streamline location ($x_0$) and angle made by the initial wave vector with the spanwise direction ($\theta^i$) for the same three $t$ as in (a)-(c). The red circles in (d)-(f) denote the location of maximum $\sigma$ at the corresponding times. A movie showing the continuous temporal evolution of the plots in (d)-(f) is uploaded as a supplementary file titled `Growth_rate_Full_analysis_Re300_Ri10^{-8}_Homogeneous.mp4`.

At a later time $t = 60$ (figure 2e), the variation of $\sigma$ is qualitatively similar to that at $t = 40$, but with a thinner and weaker instability band (i.e. spanning a smaller range in $\theta^i$) at the core, and the hyperbolic instability seemingly extending further towards the vortex centre from the vortex edge. The location of maximum $\sigma$, as denoted by the red circle, has now moved to the vortex edge, allowing us to conclude that almost purely spanwise perturbations ($\theta^i \approx 0$) at the vortex edge are likely to grow fastest at $t = 60$. At the much later time of $t = 100$ (figure 2f), the qualitative structure of $\sigma$ is very similar to that at $t = 60$, and the hyperbolic instability at the vortex edge remains dominant.

To capture the evolution of the dominant instability characteristics with time, we plot the variation of the maximum (over the entire $x_0 - \theta^i$ plane) growth rate $\sigma = \sigma^*$ with $t$ in figure 3(a) (blue solid curve). Also plotted in figure 3(a) using the yellow solid curve is the evolution of the maximum growth rate $\sigma = \sigma_c^*$ associated with the vortex centre ($x_0 = 0$). At early times, $\sigma^*$ monotonically increases with $t$, and the elliptic instability at the vortex centre remains dominant till $t \approx 40$, i.e. $\sigma^* = \sigma_c^*$ for $t \leqslant 40$. For $t > 40$, the most unstable region is away from the vortex centre, as seen by the $\sigma^*$ curve being noticeably above the $\sigma_c^*$ curve. $\sigma^*$ continues to increase till $t = 60$, at which point it attains a maximum value of 0.21, before decreasing towards smaller values. At large times, $\sigma_c^*$ converges to a nearly constant value of 0.123, with the convergence occurring around $t \approx 170$. $\sigma^*$, however, continues to decrease even at large times. Noticeable oscillations with time ($t \gtrsim 50$) are observed for $\sigma_c^*$, while the oscillations in $\sigma^*$ are relatively weaker.

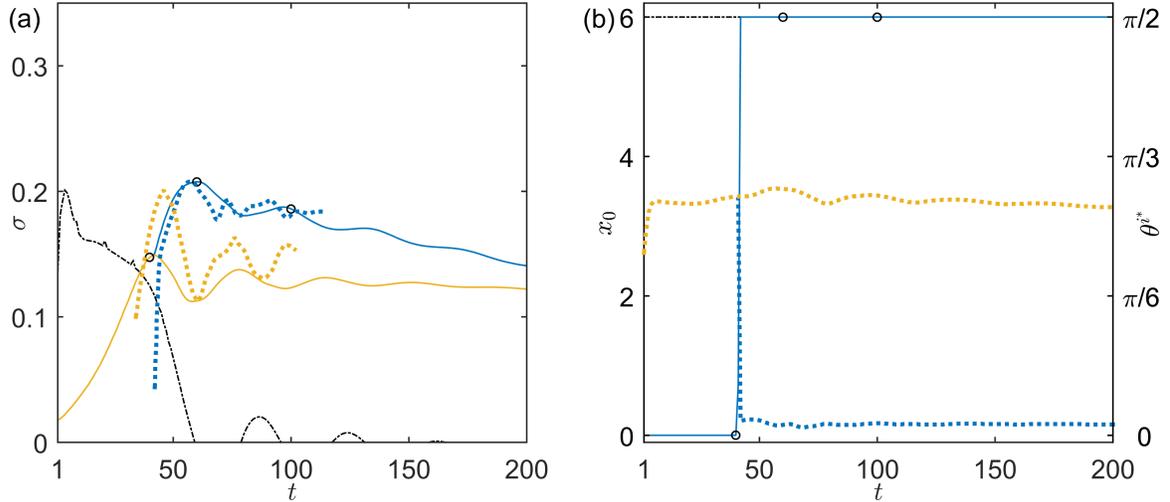

Figure 3: Dominant instability characteristics for the homogeneous case. (a) Maximum growth rate over the entire $x_0 - \theta^i$ plane, $\sigma^*$ (—), and at the centre, $\sigma_c^*$ (—), plotted as a function of time — the solid yellow curve lies on top of the solid blue curve when $\sigma^* = \sigma_c^*$ for $t \leqslant 40$. The corresponding dotted curves are the results from Klaassen & Peltier (1991) for the principal mode (····) and the central core mode (····), after appropriate vertical scaling. The black dash-dotted line indicates the variation of the growth rate associated with the primary KH instability, $\sigma_{KH}$ (-··-). (b) The most unstable streamline location $x_0^*$ (—) as an ordinate on the left axis, and the corresponding most unstable initial wave-vector angle, $\theta^{i^*}$ (····), along with the most unstable wave-vector angle at the centre, $\theta_c^{i^*}$ (····), on the right axis, plotted as a function of time. The black dash-dotted line shows the variation of location of the last closed streamline extracted from the base flow, $x_{0_l}$ (-··-), with time. The three times (∘) for which instability characteristics were presented in figures 2(d)-(f) are also shown in each panel.

Shown using the blue and yellow dotted curves in figure 3(a) are the growth rates associated with the principal mode and the central core mode, respectively, estimated by Klaassen & Peltier (1991) based on normal mode analysis. There is remarkable qualitative agreement between the viscous normal mode results of Klaassen & Peltier (1991) and our inviscid local stability estimates. We point out, however, that the growth rate estimates of Klaassen & Peltier (1991) have been multiplied by a factor of 1.5 to obtain the blue and yellow dotted curves in figure 3(a); the numerical value 1.5 results in the matching of the peak values in the blue solid and dotted curves. To quantify the extent to which the quasi-steady assumption regarding the base flow for the local stability calculations may be valid, we plot the growth rate $\sigma_{KH}$ associated with the primary KH instability. As shown by the black dashed-dotted line in figure 3(a), $\sigma_{KH}$ is smaller than $\sigma^*$ for all $t > 36$, with $\sigma_{KH}$ being very small for $t > 55$. Interestingly, $\sigma_{KH}$ displays oscillatory behaviour for $t > 55$, and its nearly in-phase relation with the oscillations in $\sigma_c^*$ suggests that the temporal evolution of the base flow causes the oscillations in $\sigma_c^*$.

In figure 3(b), we track the location of the most unstable streamline ($x_0^*$), and the corresponding most unstable wave vector ($\theta^{i^*}$) with time. As shown by $x_0^*$ plotted as the blue solid curve, the most unstable streamline remains at the centre till $t \approx 40$. Immediately afterwards, $x_0^*$ jumps to the edge of the vortex and is coincident with the outermost closed streamline at $x_0 = x_{0_l}$ extracted from the base flow. The corresponding most unstable wave vector angle $\theta^{i*}$, plotted using the blue dotted curve, is relatively close to $\pi/3$ for $t \leqslant 40$ before rapidly decreasing to $\theta^{i*} \approx 0$ for larger times. The most unstable wave vector angle $\theta_c^{i*}$ associated with the vortex centre is shown using the yellow dotted curve, and is always close to but smaller than $\pi/3$. Furthermore, we verified that $d\Gamma/d\psi > 0$ is satisfied on all the streamlines at all times, thus ruling out centrifugal instability (Sipp & Jacquin 1998) completely for $(Re, Ri) = (300, 10^{-8})$; here, $\Gamma$ is the anticlockwise circulation calculated on the streamline with stream function value $\psi$. In summary, while the vortex centre is always susceptible to elliptic instability, the dominant instability mode switches from elliptic at the vortex centre to hyperbolic at the vortex edge at $t \approx 40$.

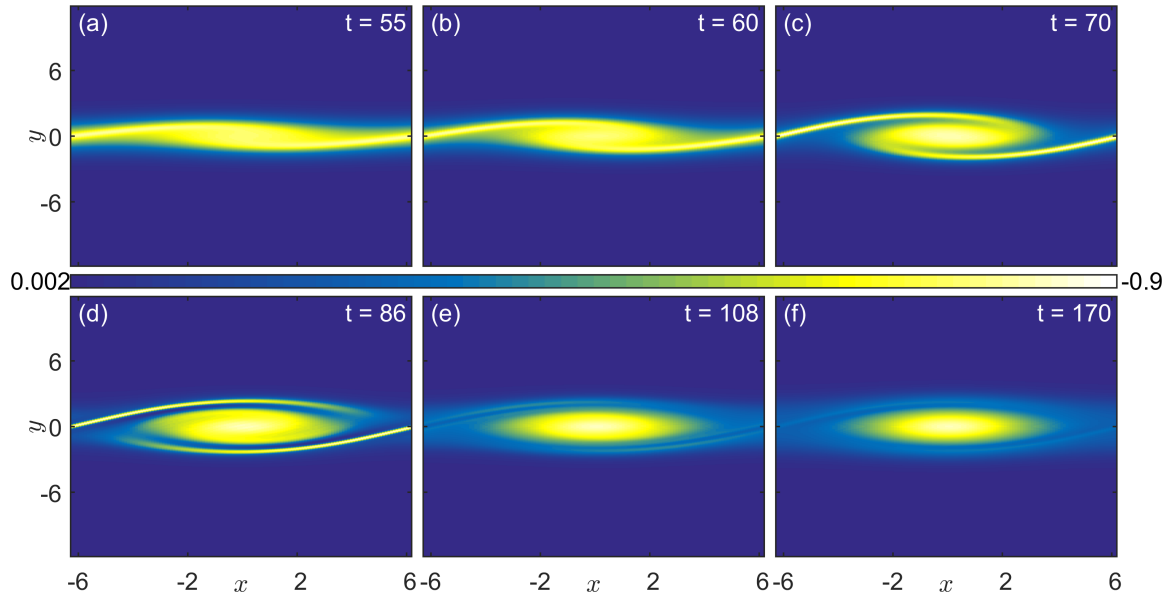

Figure 4: Vorticity field, obtained using two-dimensional numerical simulations, at six different times: (a) $t = 54$, (b) $t = 60$, (c) $t = 70$, (d) $t = 86$, (e) $t = 108$ and (f) $t = 170$ for the stratified case of $(Re, Ri) = (300, 0.08)$ (§ 3.2).

### 3.2. *Stratified flow -* $(Re, Ri) = (300, 0.08)$

In this subsection, we investigate the effects of stratification by increasing $Ri$ to 0.08, but retaining $Re$ (=300) at the same value as in § 3.1. The vorticity field, as calculated by numerical simulations, for the stratified case of $(Re, Ri) = (300, 0.08)$ at six different times are shown in figure 4. Rolling up of the shear layer, along with entrainment of the surrounding fluid into the vortex core, leads to redistribution of the vorticity in the original shear layer to both the vortex core and the braid regions. The vortex grows in size during its initial stages of rolling up, and later saturates in size at $t \approx 70$ and in shape at $t \approx 130$. Distinctive regions of anticlockwise vorticity are seen inside of the braids at $t = 86$, and it remains to be seen what their influence on the instability characteristics may be. Although the vortex geometry saturates at $t \approx 130$, the vorticity field continues to evolve with time even at larger times. We consider these simulated flow fields to be frozen at each instant and use them as steady base flows to extract corresponding closed streamlines, and subsequently solve the local stability equations on them.

At $t = 55$ (figure 5a), the elliptic instability at the vortex core is dominant, with its corresponding maximum $\sigma$ (= 0.155) occurring at $\theta^i = 49.5°$. It is noteworthy that the growth rate magnitude of this elliptic instability is comparable to its counterpart in the homogeneous case (figure 2d), suggesting that the stratification does not quantitatively influence the elliptic instability characteristics too significantly at early times. The hyperbolic instability, however, seems strongly affected by stratification, as seen by the multiple bands of instability near the vortex edge in figure 5(a). We recall from figure 2(d) that no such complex features were present for the homogeneous case. Earlier studies by Godeferd *et al.* (2001) have reported a similar effect that background rotation has on the hyperbolic instability in Stuart vortices. In our study, the in-plane buoyancy variations seem to be playing the role of spatially varying Coriolis forces that result from the background rotation.

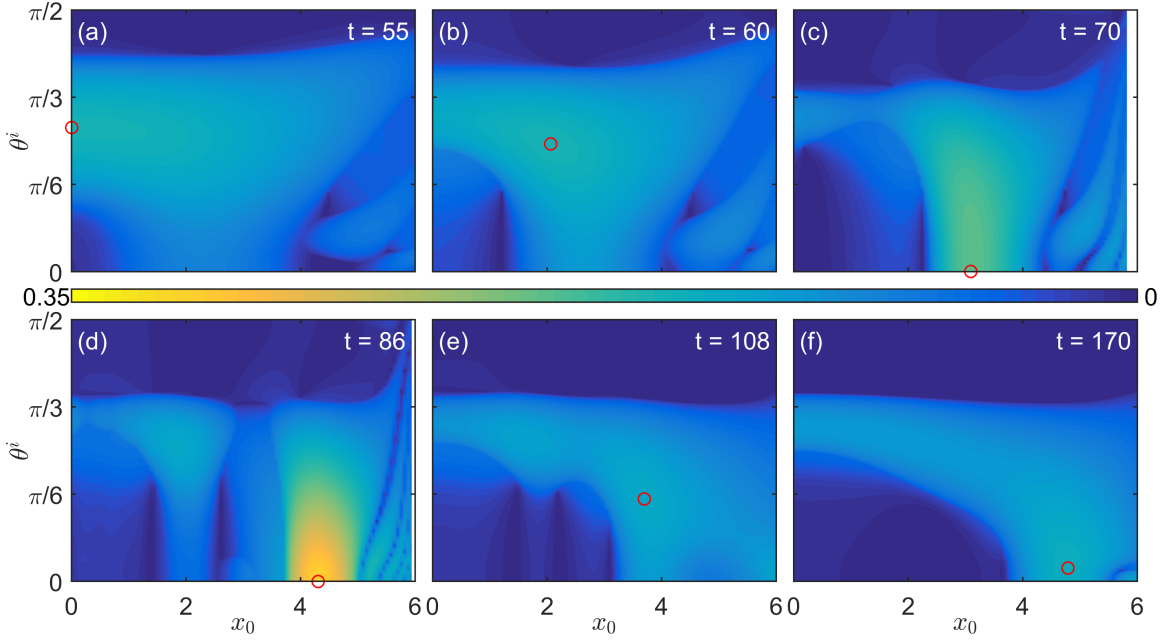

Figure 5: Temporal evolution of instability characteristics for $(Re, Ri) = (300, 0.08)$. Growth rate $\sigma$ (from the local stability approach) plotted as a function of the streamline location $x_0$ and the initial perturbation wave vector angle $\theta^i$ at the same six different times as in figure 4. $(x_0, \theta^i)$ corresponding to global maximum $\sigma$ ($\circ$) are also shown in each panel. White regions in (c) & (d) correspond to large $x_0$ at which closed streamlines do not exist. A movie showing the continuous temporal evolution of the plots in (a)-(f) is uploaded as a supplementary file titled `Growth_rate_Full_analysis_Re300_Ri0.08.mp4`.

Another distinguishing feature at $t = 55$ (figure 5a) that is not present in the homogeneous case is the occurrence of instability at $\theta^i = 0$ for the streamlines in and around the vortex core. This instability branch, that seems distinct from both the elliptic and hyperbolic branches of instability, moves away from the vortex centre with time and is characterized by its corresponding $\theta^{i*}$ hovering around zero. We henceforth refer to this new instability as the convective branch, owing to its relation with statically unstable layers, as discussed later in § 3.2.2. At $t = 60$ (figure 5b), while the elliptic instability at the core and the hyperbolic instability at the edge are still present, an intermediate streamline which is far from both the vortex core and the edge is the most unstable. The most unstable streamline at $x_0 = 2.06$ is simultaneously affected by the elliptic instability of the core and the convective branch that has moved away from the core, owing to which the corresponding most unstable $\theta^i$ ($= 44.0°$) has moved away from $\pi/3$. At a later time of $t = 70$ (figure 5c), the convective branch is the most dominant, with the corresponding $(x_0^*, \theta^{i*}) = (3.09, 0)$. This convective branch moves further to the right with time, and is strongest at $x_0^* = 4.30$ with the corresponding $\theta^{i^*}$ again being zero at $t = 86$ (figure 5d). Furthermore, at $t = 86$, we observe a second convective branch of instability centred around $x_0 = 2.06$. At a later time of $t = 108$ (figure 5e), this second convective branch has become the most dominant, while the first convective branch seems to have moved to the hyperbolic instability region. At $t = 108$, we also observe a third convective branch centred around $x_0 = 1.82$. The elliptic instability at the centre influences the second convective branch strongly enough to move the most unstable $\theta^i$ to around $\theta^{i^*} = 28.3°$. At large times ($t = 170$ shown in figure 5f), the multiple convective branches, which emerged at the vortex core at earlier times and then moved away, have coalesced into a single convective instability region with $x_0^* = 4.79$. In summary, the steady-state instability characteristics contain an elliptic branch at the core, a hyperbolic branch at the edge, and the most dominant convective branch close to but inside the periphery of the vortex.

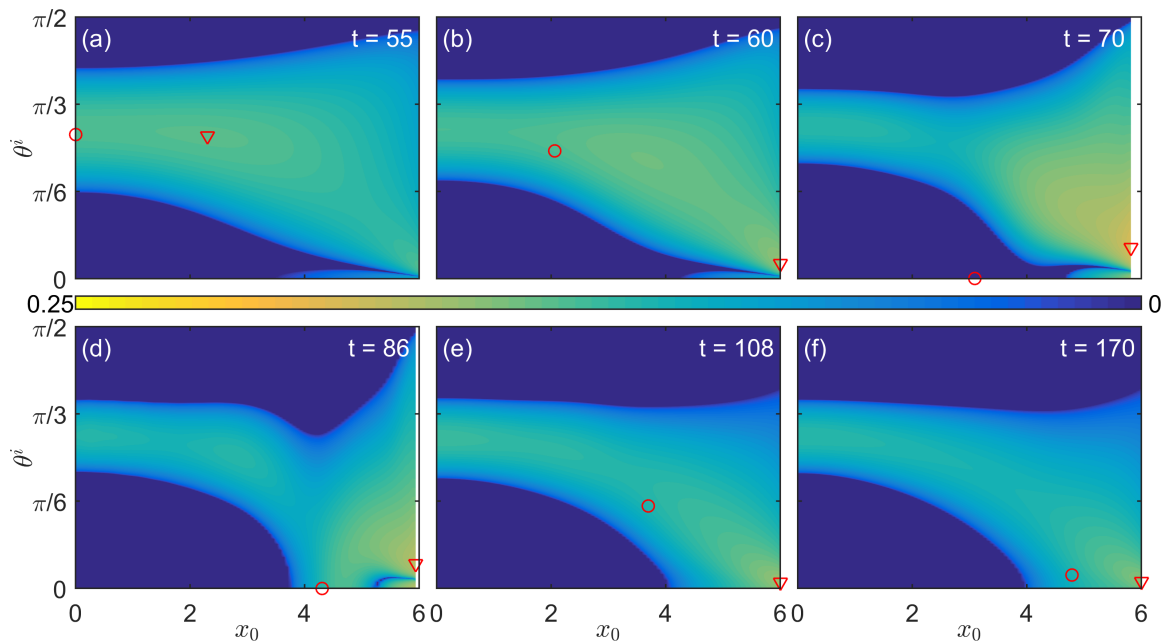

Figure 6: Temporal evolution of instability characteristics for the velocity-gradient-only (VG) analysis for $(Re, Ri) = (300, 0.08)$. Growth rate $(\sigma_{VG})$ as a function of streamline location $(x_0)$ and the initial perturbation wave-vector angle $(\theta^i)$ at the same six different times as in figures 4 & 5. $(x_0, \theta^i)$ corresponding to global maximum $\sigma$ from the full analysis (○) and $\sigma_{VG}$ (▽) are also shown in each panel. White regions in (c) & (d) correspond to large $x_0$ at which closed streamlines do not exist. A movie showing the continuous temporal evolution of the plots in (a)-(f) is uploaded as a supplementary file titled `Growth_rate_VG_analysis_Re300_Ri0.08.mp4`.

### 3.2.1. *Compartmentalized analyses*

To investigate the role of buoyancy in the various instabilities reported in figure 5, we perform a compartmentalized study in which the base flow velocity gradient or the buoyancy gradient terms are omitted from the amplitude evolution equations (2.7)-(2.8). Specifically, for the velocity-gradient-only (henceforth referred to as VG) analysis, we substitute $\nabla b_B = \mathbf{0}$, whereas in the buoyancy-gradient-only (henceforth BG) analysis, we substitute $\nabla \mathbf{u_B} = \mathbf{0}$. Both the analyses, however, consider the same streamlines, wave vectors and their evolution as in the full analysis presented in figure 5. It is noteworthy that the results from the full analysis in figure 5 are not a linear superposition of the results from the two compartmentalised analyses.

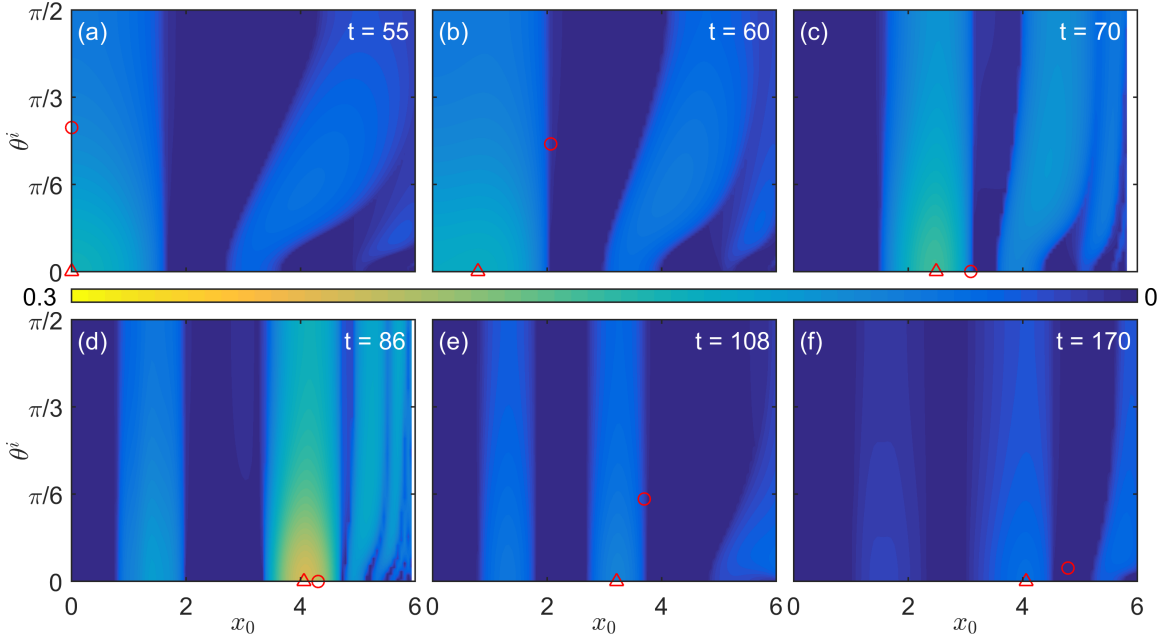

Figure 7: Temporal evolution of instability characteristics for the buoyancy-gradient-only (BG) analysis for $(Re, Ri) = (300, 0.08)$. Growth rate ($\sigma_{BG}$) as a function of streamline location ($x_0$) and the initial perturbation wave-vector angle ($\theta^i$) at the same six different times as in figures 4 & 5. $(x_0, \theta^i)$ corresponding to global maximum $\sigma$ from the full analysis ($\circ$) and $\sigma_{BG}$ ($\triangle$) are also shown in each panel. White regions in (c) & (d) correspond to large $x_0$ at which closed streamlines do not exist. A movie showing the continuous temporal evolution of the plots in (a)-(f) is uploaded as a supplementary file titled `Growth_rate_BG_analysis_Re300_Ri0.08.mp4`.

Figures 6(a)-(f) show the distribution of the growth rate $\sigma_{VG}$ from the VG analysis at the same six different times as in figure 5. The overall evolution of $\sigma_{VG}$ has strong similarities with that of $\sigma$ in the homogeneous case shown in figures 2(d)-(f). Specifically, the distribution of $\sigma_{VG}$ contains only the elliptic and hyperbolic instability branches at all times, much like what we observe for the homogeneous flow. The location of maximum $\sigma_{VG}$ (denoted by the inverted triangles) is noticeably far from the most unstable region in the full analysis (denoted by the circles) at all times shown in figure 6. There are in fact times (figure 6c) at which the location of maximum growth rate from the full analysis is stable in the VG analysis. In other words, there is no instability in $\sigma_{VG}$ in regions where the convective instability is separate from the elliptic/hyperbolic instability regions in the full analysis (figure 5c). This suggests that stratification plays a significant role in the dominant instability mechanisms observed in the full analysis.

Figures 7(a)-(f) show the distribution of the growth rate $\sigma_{BG}$ from the BG analysis at the same six different times as in figures 5 & 6. At $t = 55$, we observe the convective instability occurring at the vortex core with the corresponding $\theta^{i^*} = 0$ (vertical band of instability around the origin in figure 7a). We recall from the full analysis that the vortex core contains both the elliptic instability and the convective instability at $t = 55$, with the elliptic branch being dominant (figure 5a). Apart from the convective instability at the vortex core, $\sigma_{BG}$ also contains further instability regions closer to the vortex edge (figure 7a), which are somewhat reminiscent of the hyperbolic instability features

*Local stability analysis of homogeneous and stratified Kelvin-Helmholtz vortices* 13

in the full analysis (figure 5a). As in the full analysis, the convective instability region moves away from the core in $\sigma_{BG}$ as well ($t = 60$, figure 7b). At $t = 70$, the convective instability in $\sigma_{BG}$ has moved further to the right (figure 7c), and is not far from the dominant convective instability region from the full analysis. At $t = 86$ & $108$, the dominant convective instabilities from the BG and the full analyses occur on nearby streamlines in the same region, as indicated by the triangles and circles in figures 7(d) & (e). Furthermore, the other convective instability regions that have emerged from the core and trail behind the dominant convective instability region are also evident in figures 7(d) & (e). The complex hyperbolic instability regions continue to be present in figures 7(d) & (e), but having moved significantly towards the vortex edge. At $t = 170$, the convective instability region centered around $x_0 = 4.06$ is most dominant (figure 7f), and noticeably overlaps with the dominant convective instability in the full analysis (figure 5f). It is also noted that the instability at the vortex edge at $t = 170$ in the BG analysis is quite strong too (figure 7f). In summary, at large times, while the stratification effects are an essential ingredient of the dominant convective instability, the velocity gradients alter its location and strength. Also, as discussed later in § 3.2.2, the convective instability branches in the BG and full analyses correlate well with the number of statically unstable regions that emerge.

Figure 8 shows the evolution of the dominant instability characteristics with time from the full analysis (column 1), VG analysis (column 2) and the BG analysis (column 3) for $(Re, Ri) = (300, 0.08)$. At early times ($t \leqslant 55$), the elliptic instability at the core is the dominant instability, as is evident from $\sigma^* = \sigma_c^*$ (figure 8a) and the corresponding $\theta^{i^*}$ hovering around $\pi/3$ (figure 8d). For $t > 55$, the location $x_0^*$ moves away from the centre, with the corresponding $\theta^{i^*}$ decreasing till it becomes zero at $t = 70$. $x_0^*$ increases till $t = 107$, at which point $x_0^* = x_{0_l} = 6.18$, i.e. a region where the first convective branch and the hyperbolic-like instability at the vortex edge overlap (figure 8d). At $t = 108$, the dominant instability suddenly moves to $(x_0^*, \theta^{i^*}) = (4, 28.3°)$, which is consistent with the second convective instability branch becoming dominant, though it is influenced by the elliptic instability at the centre (figure 5e). The second convective instability branch remains dominant and moves away from the centre till $t \approx 150$, after which all subsequent convective branches coalesce for the most unstable streamline to remain around $x_0^* \approx 4.85$ at large times. The emergence and outward propagation of multiple convective instability branches from the centre are captured well by the oscillations in $\sigma_c^*$ (figure 8a); we remind ourselves that $\sigma_c^*$ is a combination of elliptic and convective instabilities at the core, as a result of which $\theta_c^{i^*}$ lies in the range $0 < \theta_c^{i^*} < \pi/3$. $x_0^*$ for all $t > 70$ is far from both the vortex core and the edge, thus confirming that the convective instability remains dominant though the elliptic instability (at the core) and the hyperbolic instability (at the edge) are still present (figure 8d).

In figure 8(a), the dotted lines represent the growth rates associated with the principal mode (blue) and the central core mode (yellow) estimated by Klaassen & Peltier (1991) using the normal mode approach. The results of Klaassen & Peltier (1991) have been scaled up by a factor 1.4 to match the peak in $\sigma^*$. There is remarkable qualitative agreement between the estimates from the local and normal mode approaches, reaffirming the relevance of the local stability approach. Quantitative differences between the two approaches, which are present in the homogeneous case as well, may be attributed to (i) absence of viscous effects in the local analysis, and (ii) not accounting for the global spatial structure of the base flow in the local approach.

The presence of a strong peak in $\sigma^*$ at $t = 86$ (figure 8a) is clearly an effect of stratification as no such feature is present in the corresponding homogeneous case (figure 3a). Also, the oscillations in $\sigma_c^*$ in the stratified case (figure 8a) are of a higher amplitude and frequency than the homogeneous case, again pointing towards the important role of stratification. To delve further into the role of stratification on the various instabilities, we present the VG and BG analyses in columns 2 and 3 of figure 8.

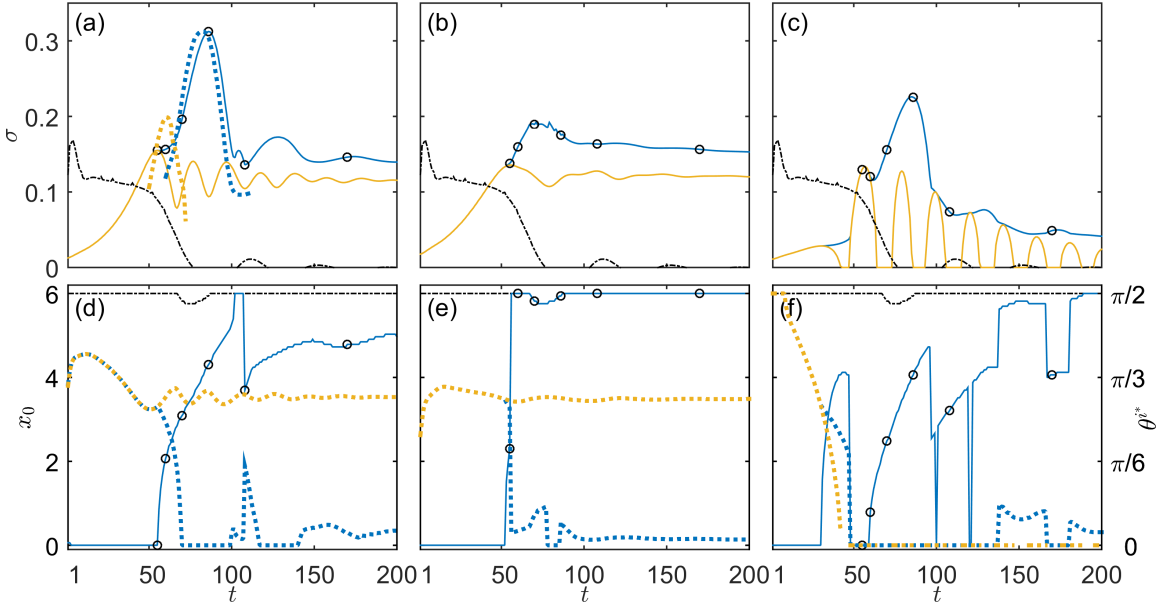

Figure 8: Dominant instability characteristics from the full analysis (column 1), the VG analysis (column 2) and the BG analysis (column 3) for $(Re, Ri) = (300, 0.08)$. (Top row) Maximum growth rate over the entire $x_0 - \theta^i$ plane, $\sigma^*$ (—), and at the centre, $\sigma_c^*$ (—), plotted as a function of time for the respective analyses — note that the solid yellow curve lies on top of the blue when $\sigma^* = \sigma_c^*$ ($t \lesssim 55$). The corresponding dotted curves in (a) are the results from Klaassen & Peltier (1991) for the principal mode (····) and the central core mode (····), after appropriate vertical scaling. The black dash-dotted lines in (a)-(c) indicate the variation of the growth rate associated with the primary KH instability, $\sigma_{KH}$ (-··-). (Bottom row) The most unstable streamline location $x_0^*$ (—) as an ordinate on the left, and the corresponding most unstable initial wave-vector angle, $\theta^{i*}$ (····), along with the most unstable wave-vector angle at the centre, $\theta_c^{i*}$ (····), on the right, plotted as a function of time. The black dash-dotted lines in (d)-(f) show the variation of location of the last closed streamline extracted from the base flow, $x_{0_l}$ (-··-), with time. The six times for which instability characteristics were presented in figures 5-7 are indicated as circles (○) in each panel.

The variation of $\sigma_{VG}^*$ in figure 8(b) closely resembles the variation of $\sigma^*$ in the homogeneous case (figure 3a). The elliptic instability at the core is dominant until $t = 52$ in figure 8(b), thus capturing the variation of $\sigma^*$ at early times in figure 8(a). Unlike in the full analysis, the dominant mode in the VG analysis switches rapidly to the hyperbolic instability at the edge of the vortex at $t \approx 56$, and remains so at large times. In contrast, we recall from figure 8(a) that the dominant mode in the full analysis changes gradually (over $55 < t < 70$) to the convective instability branch at intermediate streamlines. In summary, $\sigma_{VG}^*$ does not capture the dominant instability characteristics of the full analysis for both intermediate and large times. The BG analysis, shown in figures 8(c) & (f), reasonably captures the dominant instability characteristics in the full analysis in the interval $55 \leqslant t \leqslant 124$. Specifically, many features including (i) the main peak in $\sigma^*$ at $t = 86$, (ii) the oscillations in $\sigma_c^*$ resulting from the emergence and outward motion of convective instability branches, (iii) the outward motion of the dominant convective instability branch during $55 < t < 70$, and (iv) the switching over of the dominant mode from the first to the second convective instability branch, are all present both in the full analysis (figures 8a & d) and the BG analysis (figures 8c & f). At large times, the dominant instability regions from both the full and BG analyses are in reasonable agreement. Furthermore, both the full analysis and the BG analysis show that the instability at the vortex edge is of comparable strength to the dominant convective instability at intermediate streamlines at large times. This suggests that the velocity gradients play the catalytic role of switching the dominant mode from the instability at the vortex edge in the BG analysis to the convective branch inside the vortex periphery in the full analysis.

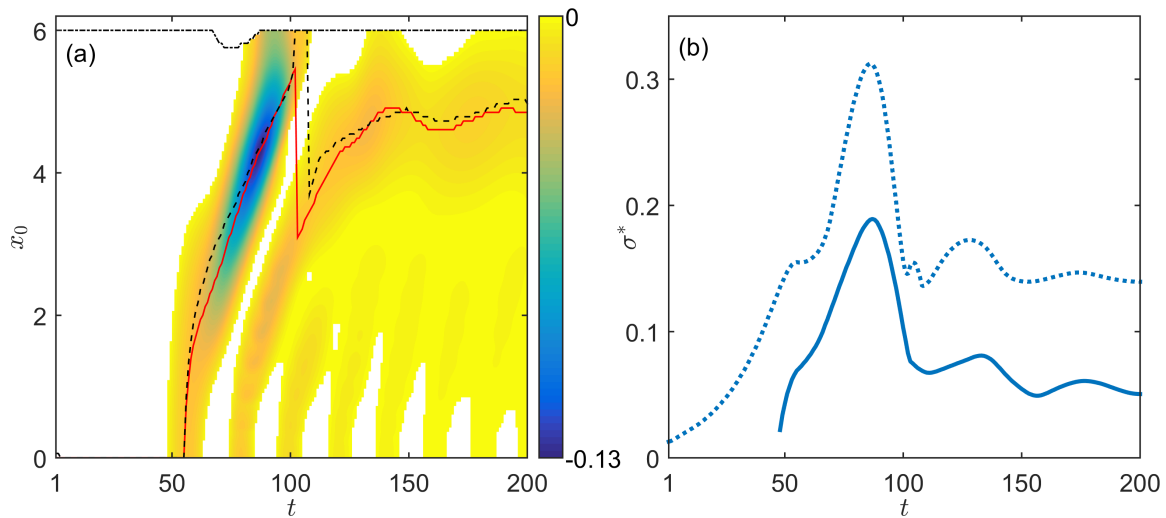

Figure 9: Statically unstable regions for $(Re, Ri) = (300, 0.08)$. (a) Minimum value of instantaneous vertical gradient of buoyancy $(\partial b_B/\partial y)$ along each streamline plotted as a function of time $t$ and the streamline location $x_0$. The most statically unstable streamline (—) is remarkably close to the most unstable $x_0$ ($x_0^*$) from the local stability calculations (- -). $x_0$ corresponding to the last closed streamline extracted from the base flow (-·-) has also been plotted as a function of $t$. White regions correspond to streamlines that entirely pass through statically stable $(\partial b_B/\partial y > 0)$ regions. (b) Growth rate associated with the static instability, $\sigma_S^*$ (—), and growth rate from the full local stability analysis, $\sigma^*$ (····) (§ 3.2.2), plotted as a function of time, $t$.

### 3.2.2. *Statically unstable layers*

The convective instability branches, identified in the stratified case, are likely to result from statically unstable layers $(\partial b_B/\partial y < 0)$ that form in the base flow (Caulfield & Peltier 2000). To investigate this aspect further, we plot the strength of the statically unstable layers that form on a given streamline, i.e. the minimum value of $\partial b_B/\partial y$ on the streamline, as a function of time $t$ and streamline location $x_0$ in figure 9(a). White regions in figure 9(a) correspond to streamlines on which $\partial b_B/\partial y > 0$ at the corresponding time instance. Unstable layers form first at the vortex centre around $t = 48$, and subsequently move outwards in time. The generation and subsequent outward motion of statically unstable layers from the vortex centre occur at almost regular intervals, which is fully consistent with the multiple convective instability branches seen in figures 5 & 7. In fact, the number of statically unstable layers generated at the vortex centre quantitatively agrees with the number of convective instability branches that emerge from the centre in both the full and BG analyses. The streamline with the strongest statically unstable layers, shown using the red solid curve in figure 9(a), is remarkably close to $x_0^*$ (black dashed curve, the most unstable streamline in the local stability analysis) at all times after the first unstable layers form in the base flow. The relation between the convective instability branches in the local stability analysis and the statically unstable layers is further verified by the qualitative agreement between $\sigma^*$ (the maximum growth rate from the local stability analysis) and $\sigma_S^* = Ri^{1/2}\sqrt{-\min_{x_0}(\partial b_B/\partial y)}$ (growth rate in a statically unstable layer with buoyancy gradient $\min_{x_0}(\partial b_B/\partial y)$) in figure 9(b). In summary, streamlines with the most severe statically unstable buoyancy gradients are good indicators of the regions of strongest convective instability at $(Re, Ri) = (300, 0.08)$.

## 4. Discussion & Conclusions

In this paper, we have performed an inviscid, three-dimensional, short-wavelength stability analysis on the Kelvin-Helmholtz vortices generated in homogeneous and stratified shear flows. The base flows were generated by numerical simulations of incompressible, viscous, two-dimensional Navier-Stokes equations within the Boussinesq approximation; the governing non-dimensional parameters are the Reynolds number $Re$ and the Richardson number $Ri$, with the Prandtl number fixed at $Pr = 1$. Assuming a quasi-steady base flow at every instance of time, growth rates as a function of the wave vector orientation $\theta^i$ (angle made with the spanwise direction) were computed for every closed streamline. For both the homogeneous ($Re = 300$, $Ri = 10^{-8}$) and stratified ($Re = 300$, $Ri = 0.08$) cases considered, the local stability analysis showed excellent qualitative agreement with the temporal evolution of the growth rates associated with the principal mode and the central core mode obtained using the global approach in Klaassen & Peltier (1991).

One of the main advantages of the computationally inexpensive local stability approach is its capability to identify mechanisms and specific streamlines associated with various instabilities. Additionally, we also performed compartmentalized analyses to isolate the roles of velocity gradients and the buoyancy gradients on the secondary instabilities. In the homogeneous flow, the elliptic instability at the core is dominant at early times when the flow field is still evolving rapidly, and the corresponding most unstable perturbations are strongly three-dimensional with an oblique wave vector. At intermediate and large times, the hyperbolic instability at the vortex edge, i.e. the streamlines passing through the neighbourhood of the hyperbolic point between consecutive KH vortices, is dominant though the elliptic instability at the core is always present. The most unstable perturbations associated with the elliptic and hyperbolic instabilities are both three-dimensional, with the corresponding wave vectors being obliquely-aligned and closely aligned with the spanwise direction, respectively.

In the stratified case of $(Re, Ri) = (300, 0.08)$, the early time instabilities are similar to the homogeneous case, but with the hyperbolic instability at the vortex edge containing complex variations with the streamline location and the wave vector angle. Additionally, a new branch of instability with relatively weak growth rates emerges at the vortex core, with the corresponding most unstable wave vector being purely spanwise. This new branch of instability, referred to as the convective branch, moves away from the vortex core, and becomes dominant soon after the primary KH instability saturates. Multiple convective instability branches get generated and then move away from the vortex core with time, and the most unstable region is sometimes observed to switch from one convective branch to another. At large times, the convective instability in a region inside the vortex periphery is dominant, while the elliptic instability at the core and the hyperbolic instability at the vortex edge are still present.

The compartmentalized analyses, i.e. including only the base flow velocity gradient (VG analysis) or the buoyancy gradient (BG analysis) terms in the amplitude evolution equations, provide further insight into the role of shear and stratification on the various instabilities in the stratified case. The dominant elliptic instability at early times in the full analysis are captured by the VG analysis, the results of which are similar to the homogeneous case analyzed in § 3.1. The VG analysis also captures the hyperbolic instability at the vortex edge, albeit with noticeably different strength and structure from the full analysis. Furthermore, the VG analysis does not capture the convective instability branches at any time, thus highlighting the role of stratification. In contrast, the BG analysis captures the multiple convective instability branches that get generated and move away from the vortex core, but does not contain any signature of the elliptic instability. The dominant convective instability in the full analysis is captured by the BG analysis at intermediate and large times. The hyperbolic instability was found to be strongly influenced by the stratification in the full analysis, which was corroborated by the similarity in the qualitative structure of the growth rate distribution close to the vortex edge between the full and BG analyses. Our observation that the evolution of secondary instabilities in a stratified shear flow is dominated by effects of velocity shear at initial times, and by effects of stratification later is consistent with the conclusions from energy budget analyses in Klaassen & Peltier (1991) and Caulfield & Peltier (2000).

The dominant convective instability regions in the full analysis were then shown to be strongly correlated with the streamlines that contain the strongest statically unstable buoyancy gradients. Furthermore, temporal variation of the convective instability growth rates in the full analysis is qualitatively described well by a simple static instability growth rate expression based on the minimum buoyancy gradient in the entire domain. In summary, at $(Re, Ri) = (300, 0.08)$, statically unstable regions in the base flow describe well the dominant short-wavelength, inviscid secondary instability characteristics at intermediate and large times.

We conclude by showing the contrasting short-wavelength secondary instability features between the homogeneous and stratified cases in figure 10. As shown in figures 10(a)-(c), which correspond to the homogeneous case, the elliptic instability at the vortex core dominates at early times (figure 10a), before being taken over by the hyperbolic instability at the vortex edge at later times (figures 10b-c). In the stratified case, the spatial distribution of the growth rate at $t = 55$ (figure 10d) bears resemblances with the early time behavior in the homogeneous case, with the elliptic instability at the core being dominant. At $t = 60, 70$ (figures 10e-f), a distinct dominant convective instability occurs on streamlines that are intermediate to both the vortex centre and edge; such a feature is completely absent in the homogeneous case. The convective instability region then moves away from the vortex centre, as additional convective instability branches emerge from the centre (figures 10g-h). At $t = 108$, the first convective instability branch seems to have moved to the vortex edge, while two other convective instability branches are observed at intermediate streamlines. All the convective instability branches except the first coalesce into a single convective instability branch inside the periphery of the vortex at large times (figure 10i).

In conclusion, we have established the local stability approach as a powerful framework that can be used to understand secondary instabilities in stratified shear flows. It would be worthwhile to investigate the specific roles of the various instabilities and unstable regions, identified using the local stability approach, in the formation of various spatial and temporal features as the flow transitions to turbulence. A systematic study (using the local stability approach) to understand the effects of $Re$, $Ri$, $Pr$, the ratio between the initial shear and buoyancy layer widths, and the far stream boundaries would not be too computationally prohibitive, and can be potentially insightful. Our results have also motivated a study on the effects of in-plane buoyancy gradients on previously known instabilities in idealized vortex models. Finally, the role of background rotation could be investigated to move closer to understanding secondary instabilities in atmospheric and oceanic boundary layer flows.

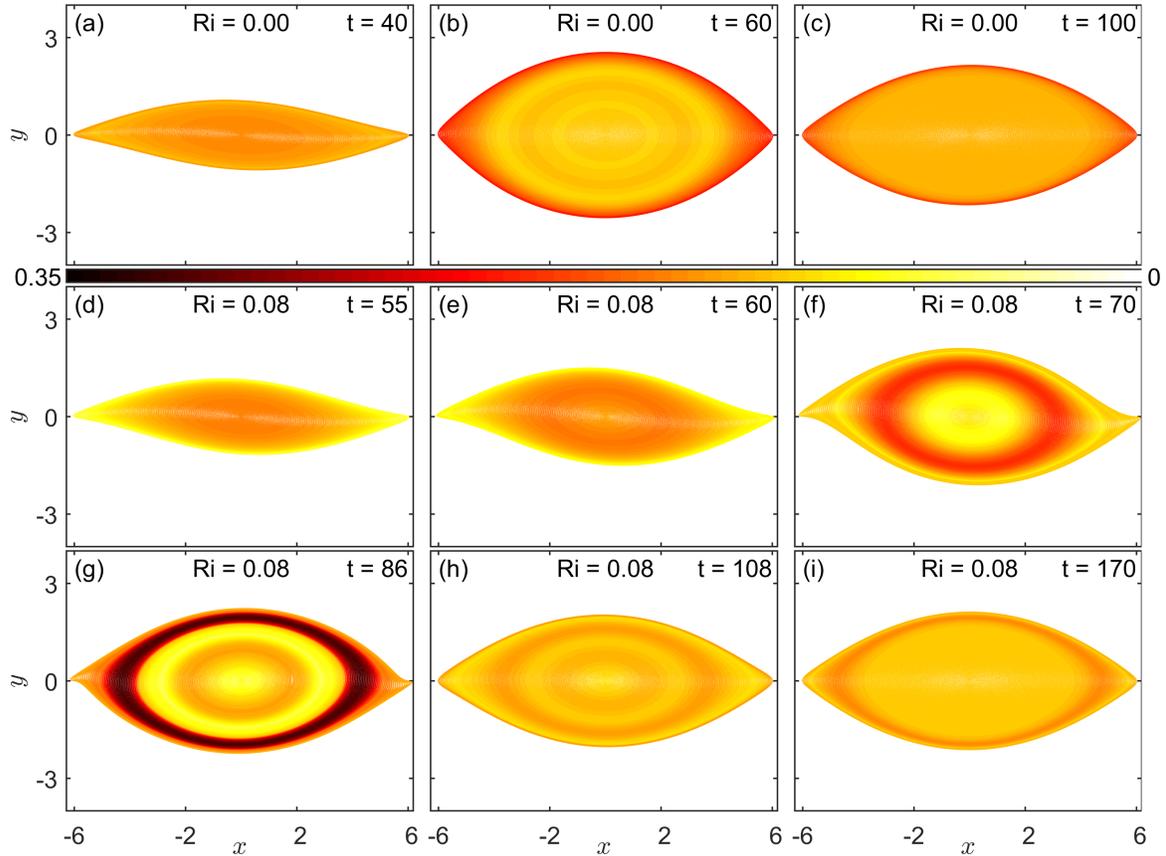

Figure 10: Streamlines at representative times (shown within each plot) plotted in a colour that indicates the maximum growth rate on them at the corresponding time. (a)-(c) The homogeneous case of $(Re, Ri) = (300, 10^{-8})$, (d)-(i) the stratified case of $(Re, Ri) = (300, 0.08)$. A movie showing the continuous temporal evolution of the plots for the homogeneous case in (a)-(c) is uploaded as a supplementary file titled `Growth_rate_streamlines_Re300_Ri10^{-8}_Homogeneous.mp4`. The corresponding movie for the stratified case in (d)-(i) is uploaded as a supplementary file titled `Growth_rate_streamlines_Re300_Ri0.08.mp4`.

## Acknowledgements

H.M.A. thanks Charpak fellowship, Embassy of France in India for funding his visit to École Polytechnique, Palaiseau, France during this project. The authors also thank the Ministry of Earth Sciences, Government of India for financial support under the Monsoon Mission Grant MM/2014/IND-002.